\documentstyle[12pt, psfig]{iopart}

\begin{document}

\pagenumbering{arabic}

\title{Searching for New Physics with Ultrahigh Energy Cosmic Rays}

\author{Floyd       W      Stecker\footnote{Corresponding      author:
Floyd.W.Stecker@nasa.gov}}

\address{Astrophysics  Science Division\\  NASA  Goddard Space  Flight
Center, Greenbelt, MD 20771, USA}

\author{Sean T Scully}

\address{Dept.  of  Physics and Astronomy\\  James Madison University,
Harrisonburg, VA 22807, USA}

\vskip 12pt

\begin{abstract}

Ultrahigh energy  cosmic rays that produce giant  extensive showers of
charged  particles  and photons  when  they  interact  in the  Earth's
atmosphere  provide a  unique  tool  to search  for  new physics.   Of
particular  interest is  the  possibility of  detecting  a very  small
violation  of  Lorentz  invariance  such  as may  be  related  to  the
structure of space-time near the  Planck scale of $\sim 10^{-35}$m. We
discuss here the possible signature of Lorentz invariance violation on
the spectrum of ultrahigh energy  cosmic rays as compared with present
observations of giant air  showers.  We also discuss the possibilities
of using  more sensitive detection techniques to  improve searches for
Lorentz invariance violation in the future. Using the latest data from
the  Pierre  Auger  Observatory, we  derive  a  best  fit to  the  LIV
parameter of $3.0^{+1.5}_{-3.0}  \times 10^{-23}$, corresponding to an
upper limit  of $4.5  \times 10^{-23}$ at  a proton Lorentz  factor of
$\sim 2 \times 10^{11}$.  This result has fundamental implications for
quantum gravity models.

\end{abstract}

\section{Introduction}

\subsection{Why Test Fundamental Physics at Ultrahigh Energies?} 

Owing to the uncertainty principle, it has long been realized that the
higher the particle energy attained,  the smaller the scale of physics
that can  be probed. Thus, optical,  UV and X-ray  observations led to
the  understanding   of  the  structure  of   the  atom,  $\gamma$-ray
observations led  to an understanding  of the structure of  the atomic
nucleus, and  deep inelastic  scattering experiments with  high energy
electrons  led to  an understanding  of the  structure of  the proton.
Accelerator  experiments  have  led  to an  understanding  of  quantum
chromodynamics  and  it  is  hoped  that  the  Large  Hadron  Collider
\cite{ev07} will eventually reveal new  physics at the TeV scale. This
could  lead  to  the  discovery  of  the  predicted  Higgs  boson  and
supersymmetric particles.  To go much beyond this scale of fundamental
physics, to search  for clues to a grand  unification theory, and even
Planck  scale physics,  one must  turn  to the  extreme high  energies
provided by the  cosmic generators with which Nature  has provided us.
In this focus paper, we  will concentrate on searching for conjectured
ultrahigh energy modifications of special relativity. This search will
be  based on  obtaining data  on the  spectrum of  cosmic rays  at the
highest energies  observed and even  beyond, using present  and future
detection techniques.

\subsection{Theoretical Motivation for High Energy Violation of 
Lorentz Invariance}

The theory of relativity is, of course, one of the fundamental pillars
of modern  physics.  However, because of the  problems associated with
merging relativity  with quantum  theory, it has  long been  felt that
relativity will have to be modified  in some way in order to construct
a quantum theory of gravitation.

The group of Lorentz  transformations delineated by special relativity
can be described as a  high energy modification of the unbounded group
of Galilean transformations. Since the Lorentz group is also unbounded
at the  high boost (or high energy)  end, in principle it  may also be
subject to  modifications in  the high boost  limit.  There is  also a
fundamental relationship between  the Lorentz transformation group and
the  assumption  that space-time  is  scale-free,  since  there is  no
fundamental length  scale associated with the  Lorentz group. However,
as noted  by Planck ~\cite{pl99},  there is a  potentially fundamental
scale associated  with gravity, {\it  viz.}, the Planck  scale.  Thus,
there  has  been a  particular  interest  in  the possibility  that  a
breakdown of Lorentz invariance (LI) may be associated with the Planck
scale, $\lambda_{Pl}  = \sqrt{G\hbar /c^3} \sim 10^{-35}$  m, owing to
various speculations regarding  quantum gravity scenarios.  This scale
corresponds  to  an  energy  (mass)   scale  of  $M_{Pl}  =  \hbar  c/
\lambda_{Pl} \sim 10^{19}$ GeV.

It is at the Planck scale where quantum effects are expected to play a
key  role  in determining  the  effective  nature  of space-time  that
emerges in the classical continuum limit.  The idea that LI may indeed
be only  approximate has  been explored within  the context of  a wide
variety  of  suggested  new  Planck-scale  physics  scenarios.   These
include  the concepts  of deformed  relativity, loop  quantum gravity,
non-commutative  geometry, spin  foam models,  and some  string theory
models.    Such   theoretical    explorations   and   their   possible
consequences, such as  observable modifications in the energy-momentum
dispersion  relations  for  free  particles  and  photons,  have  been
discussed under the general heading of ``Planck scale phenomenology''.
There is  an extensive literature on this  subject.  (See ~\cite{ma05}
for  a  review;  some   recent  references  are  Refs.~\cite{el08}  --
~\cite{he09}.)

\subsection{Testing Special Relativity using Astrophysical Observations}

It has been proposed that violation of LI at a high energy such as the
Planck  scale  could have  astrophysical  consequences  that might  be
manifested  in  a suppressed  form  at  an  energy scale  $<<  M_{Pl}$
~\cite{sa72} --  ~\cite{ac98}. A surprising result  of subsequent work
has  been the  conclusion that  several potential  effects  of Lorentz
invariance violation (LIV) can be explored and tested at energies many
orders of  magnitude below  the energy of  the Planck scale  (See {\it
e.g.}, Ref. \cite{ma05,ja04} and references therein.)

Among the relevant astrophysical tests, we focus here on the ultrahigh
energy cosmic-ray  sector.  Astrophysically produced  ultrahigh energy
particles  are  the perfect  vehicles  to  explore  the potential  for
detection of  possible violations  of special relativity  at ultrahigh
energies. One  may also search  for possible evidence of  Planck scale
physics  and  quantum   gravity  through  photon  propagation  effects
~\cite{ac98},\cite{el08}.  Such effects may be revealed by space-based
observations from  the {\it  SWIFT} $\gamma$-ray burst  detector ({\tt
http://heasarc.gsfc.nasa.gov/docs/swift/swiftsc.html}),  and  the {\it
Fermi}   ({\tt    http://fermi.gsfc.nasa.gov/})   $\gamma$-ray   Space
Telescope  ~\cite{abdo09}.   We   will  concentrate  here  on  present
observations of  ultrahigh energy cosmic  rays.  We will  also discuss
future satellite  programs proposed to make  observations of ultrahigh
energy  cosmic-ray  air showers  from  space  such  as {\it  JEM-EUSO}
(Extreme  Universe  Space   Observatory)  \cite{EUSO}  and  {\it  OWL}
(Orbiting Wide-Angle Light Collectors) \cite{OWL} (See section 7.)

\section{Ultrahigh Energy Cosmic Rays}

\subsection{Extragalactic Origin of Ultrahigh Energy Cosmic Rays}

Ultrahigh  energy cosmic rays  (UHECRs) produce  giant air  showers of
charged  particles  when  they  impinge  on  the  Earth's  atmosphere.
Observational  studies of  these  showers have  been undertaken  using
scintillator  arrays and with  atmospheric fluorescence  detectors. In
this  manner the  total energies  and  atomic weights  of the  primary
particles can be determined from the shower characteristics. The total
energy of the primary incoming particle can be deduced from the number
of secondary  charged particles produced  at a fiducial  distance from
the shower axis or the  amount of atmospheric fluorescence produced by
the shower.  A  rough measurement of the atomic  weight of the primary
can  be  obtained from  the  determining  the  height of  the  initial
interaction in the atmosphere.

The  history  of UHECR  detection  goes  back  almost half  a  century
\cite{li63}.  Owing  to their  observed global isotropy  and ultrahigh
energy  that allows  them to  be unfettered  by the  galactic magnetic
field,  cosmic rays  above  10 EeV  (1  EeV $\equiv  10^{18}$ eV)  are
believed to be  of extragalactic origin. This fact,  together with the
absence  of a  correlation  of arrival  directions  with the  galactic
plane, indicates that  if protons are the primary  particles that make
up the ultrahigh  energy cosmic radiation, these protons  should be of
extragalactic origin.

The  large air  shower detector  arrays  and, in  particular the  {\it
Auger} array, ({\tt  http://www.auger.org/}) (a.k.a.  the Pierre Auger
Observatory (PAO)) have opened up two potential new areas of research.
One area is a new field  of ultrahigh energy particle astronomy -- the
identification  and  exploration  of  powerful  extragalactic  sources
capable  of accelerating  cosmic rays  to energies  above 1  joule per
particle.  The second area, which is the topic of this focus paper, is
the field  of potential new  ultrahigh energy particle physics  -- the
search  for new  physical processes  that may  occur at  energies much
greater than those produced in man-made laboratories.

\subsection{The ``GZK Effect''}

Shortly after the discovery  of the 3K cosmogenic background radiation
(CBR),  Greisen  \cite{gr66}  and  Zatsepin  and  Kuz'min  \cite{za66}
predicted that pion-producing interactions  of such cosmic ray protons
with the CBR should produce a spectral cutoff at $E \sim$ 50 EeV.  The
flux  of  ultrahigh energy  cosmic  rays  (UHECR)  is expected  to  be
attenuated by such photomeson  producing interactions.  This effect is
generally known as the ``GZK  effect''.  Owing to this effect, protons
with energies above $\sim$100~EeV  should be attenuated from distances
beyond $\sim 100$ Mpc because  they interact with the CBR photons with
a resonant photoproduction of pions \cite{st68}.

The  flux and  spectrum of  the secondary  ultrahigh  energy neutrinos
resulting  from  the  decay   of  the  photoproduced  pions  was  also
subsequently  calculated  \cite{st73,en01}.   Photons with  comparable
ultrahigh  energy  have  much  smaller  mean-free-paths  because  they
pair-produce  electrons  and   positrons  by  interacting  with  radio
background photons  and are  thus attenuated.  The  attenuation length
for photons is somewhat uncertain, because of the uncertainties in our
knowledge  of   the  flux  and   spectrum  of  the   radio  background
\cite{pr96,st03a}.

\begin{figure}
\centerline{\psfig{figure=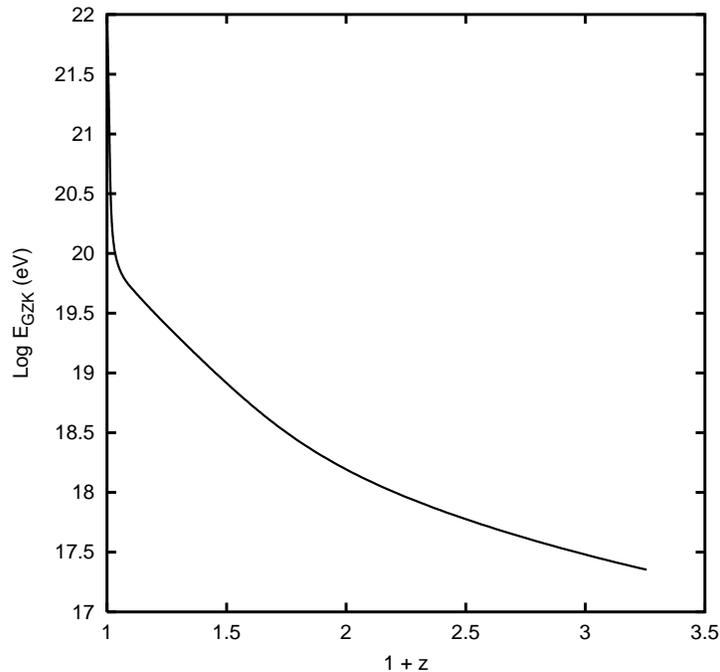,height=14cm}}
\vspace{-1.5cm}
\caption{The GZK cutoff energy, defined  as the energy predicted for a
flux decrease  of $1/e$  owing to intergalactic  photomeson production
interactions, as a function of redshift \protect \cite{sc02}.}
\label{egzk}
\end{figure}

The  GZK  effect is  not  a  true cutoff,  but  a  suppression of  the
ultrahigh  energy  cosmic  ray  flux  owing  to  an  energy  dependent
propagation time  against energy losses  by such interactions,  a time
which is only $\sim$300 Myr  for 100 EeV protons \cite{st68}.  At high
redshifts, $z$,  the target photon density increases  by $(1+z)^3$ and
both the photon and initial cosmic ray energies increase by $(1+z)$. A
plot of the  GZK energy as a function of  redshift, calculated for the
$\Lambda$CDM cosmology, is shown  in Figure \ref{egzk} \cite{sc02}. If
the source  spectrum is  hard enough, there  could also be  a relative
enhancement just  below the  ``GZK energy'' owing  to a  ``pileup'' of
cosmic rays starting out at  higher energies and crowding up in energy
space at or below the predicted GZK cutoff energy~\cite{st89}.  
At energies in the 1-10 EeV range, pair production interactions should 
take a bite out of the UHECR spectrum.

\subsection{Astrophysical Studies of Fundamental Physics at 
Ultrahigh Energies}

Some ``trans-GZK'' hadronic showers  with energies above the predicted
GZK cutoff  energy have been reportedly observed  by both scintillator
and  fluorescence detectors,  particularly by  the  scintillator array
group at Akeno \cite{ta03},  in apparent contradiction to the expected
GZK  attenuation  effect.   While  there  is less  evidence  for  such
interesting events  from fluorescence  detectors, the {\it  Fly's Eye}
fluorescence  detector  reported the  detection  of  a  320 EeV  event
\cite{bi94},  an energy  that is  a factor  of $\sim$5  above  the GZK
cutoff  energy.  The  evidence for  such trans-GZK  events has  been a
prime  motivation for  suggesting violation  of Lorentz  invariance at
ultrahigh energies \cite{sa72,ki72}.

However,  the existence of  a physically  significant number  of UHECR
events at  trans-GZK energies has recently been  called into question.
The latest {\it Auger}  data \cite{sch09},\cite{data} as well as those
from  {\it   HiRes},  \cite{ab08},  have  both   been  interpreted  as
indicating  a  GZK  cutoff.  This  has  led  many  in the  cosmic  ray
community to assume  that there is no new physics  to be discovered at
ultrahigh  energies. Thus,  the  emphasis  in the  field  has been  on
ultrahigh energy particle {\it  astronomy}, {\it i.e.}, the attempt to
determine which nearby extragalactic  objects accelerate and emit such
high energy particles. However, the  subject of this paper will be the
search for {\it new physics} at ultrahigh energies.  In particular, we
will discuss the features in  the ultrahigh energy cosmic ray spectrum
that would be a signal  of Lorentz violation and possible Planck scale
physics and would also be compatible with present observational data.

\section{Violating Lorentz Invariance - A Framework}

In this paper we will  take the phenomenological approach to exploring
the effects of LIV pioneered by Coleman and Glashow ~\cite{co99}. They
have proposed a  simple formalism {\it via} postulating  a small first
order  perturbation in the  free-particle Lagrangian.   This formalism
has  the  advantages of  (1)  simplicity,  (2)  preserving the  $SU(3)
\otimes SU(2)  \otimes U(1)$ standard model of  strong and electroweak
interactions,  (3)  having the  perturbative  term  in the  Lagrangian
consist of operators of mass  dimension four that thus preserves power
counting renormalizability, and (4)  being rotationally invariant in a
preferred frame that  can be taken to  be the rest frame of  the 2.7 K
cosmic  background  radiation\footnote{See  Ref.~\cite{alt07a}  for  a
generalization to the non-isotropic  case.}  This formalism has proven
useful   in    exploring   astrophysical   data    for   testing   LIV
~\cite{co99},\cite{sg01}.

Coleman  and   Glashow  start  with   a  standard-model  free-particle
Lagrangian,

\begin{equation}
{\cal L} = \partial_{\mu} \Psi ^ * {\bf Z} \partial^{\mu}\Psi - \Psi ^
* {\bf M}^2\Psi
\end{equation}

\noindent  where $\Psi$ is  a column  vector of  $n$ fields  with U(1)
invariance  and the  positive Hermitian  matrices ${\bf  Z}$  and {\bf
M}$^2$ can be  transformed so that ${\bf Z}$ is  the identity and {\bf
M}$^2$ is diagonalized to produce the standard theory of $n$ decoupled
free fields.

They  then add a  leading order  perturbative, Lorentz  violating term
constructed from only spatial  derivatives with rotational symmetry so
that

\begin{equation}
{{\cal  L}  \rightarrow  {\cal  L}  +  \partial_i\Psi  {\bf  \epsilon}
\partial^i\Psi} ,
\end{equation}

\noindent where  {\bf $\epsilon$} is a  dimensionless Hermitian matrix
that commutes with {\bf M}$^2$ so that the fields remain separable and
the  resulting  single particle  energy-momentum  eigenstates go  from
eigenstates  of {\bf  M}$^2$  at  low energy  to  eigenstates of  {\bf
$\epsilon$} at high energies.

To  leading order,  this  term  shifts the  poles  of the  propagator,
resulting in the free particle dispersion relation

\begin{equation}
E^2 = \vec{p} \ ^{2} + m^{2} + \epsilon \vec{p} \ ^2.
\label{dispersion}
\end{equation}

This can be put in the standard form for the dispersion relation

\begin{equation}
E^2 = \vec{p \ }{^2}c_{MAV}^2 + m^{2} c_{MAV}^4,
\end{equation}

\noindent by  shifting the  renormalized mass by  the small  amount $m
\rightarrow m/(1+\epsilon)$  and shifting the velocity from  c (=1) by
the amount $c_{MAV} = \sqrt (1 + \epsilon) \simeq 1 + \epsilon/2$.

The group velocity is given by

\begin{equation}
{{\partial  E}\over{\partial |\vec{p}|}}  = {{|\vec{p}|}  \over {\sqrt
{|\vec{p}|^2 + m^2 c_{MAV} ^2}}} c_{MAV},
\label{MAV}
\end{equation}

\noindent which goes  to $c_{MAV}$ in the limit  of large $|\vec{p}|$.
Thus,  Coleman  and  Glashow  identify  $c_{MAV}$ to  be  the  maximum
attainable velocity  of the free  particle.  Using this  formalism, it
becomes  apparent that,  in  principle, different  particles can  have
different maximum attainable velocities  (MAVs) which can be different
from $c$.  Hereafter,  we denote the MAV of a particle  of type $i$ by
$c_{i}$ and the difference

\begin{equation}
c_{i}  -  c_{j}  ~=~ {{\epsilon_{i}-\epsilon_{j}}\over{2}}~  \equiv  ~
\delta_{ij}.
\label{deltadef}
\end{equation}

There  are  other popular  formalisms  that  are  inspired by  quantum
gravity models or  by speculations on the nature  of space-time at the
Planck scale.   There are formidable obstacles to  constructing a true
quantum   gravity   theory.    Among   these   is   the   problem   of
renormalizablility  \cite{sh07}.  Largranians  involving  operators of
mass    dimension    greater    than    four   are    generally    not
renormalizable. However, in the  context of an effective field theory,
one can  postulate Lagrangians containing operators  of mass dimension
$\ge$    5    with     suppression    factors    as    multiples    of
$M_{Pl}~$\cite{ja04},\cite{mp03}.  This  leads to dispersion relations
having  a  series  of   smaller  and  smaller  terms  proportional  to
$p^{n+2}/M_{Pl}^n  \simeq  E^{n+2}/M_{Pl}^n$,  with  $n \ge  1$.   The
astrophysical implications  of this  formalism have been  discussed in
the  literature  \cite{ja04},  \cite{al00}-\cite{ma09}.   However,  in
relating LIV to the observational data  on UHECRs, it is useful to use
the simpler  formalism of Coleman  and Glashow. 
Given the limited energy range of the UHECR data relevant to the
GZK effect,  this formalism  can later be  related to  possible Planck
scale phenomena and quantum gravity models of various sorts
(See Section 6.2). 

Let us consider  the photomeson production process leading  to the GZK
effect. Near threshold, where single pion production dominates,

\begin{equation}
p + \gamma \rightarrow p + \pi.
\end{equation}

Using the  normal Lorentz  invariant kinematics, the  energy threshold
for  photomeson interactions  of UHECR  protons of  initial laboratory
energy $E$ with  low energy photons of the  CBR with laboratory energy
$\omega$, is  determined by the relativistic invariance  of the square
of  the  total  four-momentum   of  the  proton-photon  system.   This
relation, together with the threshold inelasticity relation $E_{\pi} =
m/(M  +  m)  E$  for  single pion  production,  yields  the  threshold
conditions for head on collisions in the laboratory frame

\begin{equation}
4\omega E = m(2M + m)
\end{equation}

\noindent for the proton, and

\begin{equation}
4\omega E_{\pi} = {{m^2(2M + m)} \over {M + m}}
\label{pion}
\end{equation}

\noindent in terms of the pion energy, where M is the rest mass of the
proton and m is the rest mass of the pion~\cite{st68}.

If LI  is broken so  that $c_\pi~ >~  c_p$, it follows  from equations
(\ref{dispersion}),   (\ref{deltadef})  and   (\ref{pion})   that  the
threshold energy for  photomeson is altered because the  square of the
four-momentum  is shifted  from  its  LI form  so  that the  threshold
condition in terms of  the pion energy becomes\footnote{We assume here
that protons  and pions are kinematically independent  entities. For a
treatment of these  particles as composites of quarks  and gluons, see
Ref.~\cite{ga04}.}

\begin{equation}
4\omega E_{\pi}  = {{m^2(2M +  m)} \over {M  + m}} + 2  \delta_{\pi p}
E_{\pi}^2
\label{LIVpi}
\end{equation}

\noindent 

Equation (\ref{LIVpi})  is a quadratic  equation with real  roots only
under the condition

\begin{equation}
\delta_{\pi p}  \le {{2\omega^2(M  + m)} \over  {m^2(2M +  m)}} \simeq
\omega^2/m^2.
\label{root}
\end{equation}

Defining  $\omega_0 \equiv  kT_{CBR} =  2.35 \times  10^{-4}$  eV with
$T_{CBR} = 2.725\pm 0.02$ K, equation (\ref{root}) can be rewritten

\begin{equation}
\delta_{\pi p} \le 3.23 \times 10^{-24} (\omega/\omega_0)^2.
\label{CG}
\end{equation}

\section{Kinematics}

If LIV occurs and $\delta_{\pi p} > 0$, photomeson production can only
take place for interactions of  CBR photons with energies large enough
to satisfy equation (\ref{CG}). This condition, together with equation
(\ref{LIVpi}), implies  that while photomeson  interactions leading to
GZK suppression can occur for ``lower energy'' UHE protons interacting
with higher energy CBR photons on the Wien tail of the spectrum, other
interactions involving higher energy  protons and photons with smaller
values  of  $\omega$ will  be  forbidden.   Thus,  the observed  UHECR
spectrum may  exhibit the characteristics of GZK  suppression near the
normal GZK threshold, but the UHECR spectrum can ``recover'' at higher
energies  owing to  the  possibility that  photomeson interactions  at
higher  proton energies  may be  forbidden.   We now  consider a  more
detailed quantitative  treatment of this possibility,  {\it viz.}, GZK
coexisting with LIV.

The  kinematical  relations   governing  photomeson  interactions  are
changed  in  the  presence  of  even  a  small  violation  of  Lorentz
invariance.      Following     equations    (\ref{dispersion})     and
(\ref{deltadef}), we denote

\begin{equation}
E^2=p^2+2\delta _a p^2 +{m_a}^2
\label{dispersiona}
\end{equation}

\noindent where $\delta _a$ is  the difference between the MAV for the
particle {\it a} and the speed  of light in the low momentum limit ($c
= 1$).

The square of the cms energy of particle $a$ is then given by

\begin{equation}
\sqrt{s_a} = \sqrt{E^2-p^2} = \sqrt{2\delta_a p^2 + m_a^2} \ge 0.
\label{restmass}
\end{equation}

Owing to  LIV, in the cms the  particle will not generally  be at rest
when $p = 0$ because

\begin{equation}
v = {{\partial E} \over {\partial p}} \neq {{p}\over {E}}.
\end{equation}

The modified kinematical relations containing LIV have a strong effect
on the amount of energy transfered  from a incoming proton to the pion
produced in  the subsequent interaction, {\it  i.e.}, the inelasticity
\cite{al03,ss08}.   The  total inelasticity,  $K$,  is  an average  of
$K_{\theta}$, which depends on the angle between the proton and photon
momenta, $\theta$:
\begin{equation}
K = \frac{1}{\pi }\int\limits_0^\pi {K_\theta d\theta }.
\label{Ktot}
\end{equation}

The primary  effect of LIV on  photopion production is  a reduction of
phase space allowed for the interaction.  This results from the limits
on the allowed range of interaction angles integrated over in order to
obtain  the  total   inelasticity  from  equation  (\ref{Ktot}).   For
real-root solutions for  interactions involving higher energy protons,
the  range of  kinematically allowed  angles in  equation (\ref{Ktot})
becomes severely  restricted.  The modified  inelasticity that results
is  the   key  in  determining   the  effects  of  LIV   on  photopion
production. The inelasticity rapidly  drops for higher incident proton
energies.

As  shown  in Ref.~\cite{co99},  in  order  to  modify the  effect  of
photopion production  on the  UHECR spectrum above  the GZK  energy we
must have $\delta_{\pi} > \delta_p$, {\it i.e.}, $\delta_{\pi p} > 0$.
We note that a constraint can be put on $\delta_{p\gamma}$ in the case
where $\delta_{p\gamma} > 0$ In this case, protons will have a maximum
allowed energy

\begin{equation}
E_{\rm max}= m_p\,\sqrt{1/2\delta_{p\gamma}}\;.
\end{equation} 

\noindent above  which protons traveling  faster than light  will emit
light  at  all frequencies  by  the  process  of `vacuum  \v{C}erenkov
radiation'   ~\cite{co99},~\cite{sg01},~\cite{alt07}.    This  process
occurs rapidly,  so that the  energy of the superluminal  protons will
rapidly fall back to  energy $E_{\rm max}$. Therefore, because UHECRs,
assumed here to  be protons, have been observed up  to an upper energy
of $E_{U} \simeq$ 320 EeV ~\cite{bi94}, it follows that

\begin{equation}
\delta_{p\gamma}   \le   {{m_p^2}\over{2E_{U}^2}}   \simeq  5   \times
10^{-24}.
\label{vc}
\end{equation}

Our requirement that $\delta_{\pi  p} > 0$ precludes the `quasi-vacuum
\v{C}erenkov  radiation'  of  pions,   {\it  via}  the  rapid,  strong
interaction,  pion emission process,  $p \rightarrow  N +  \pi$.  This
process would be allowed by LIV  in the case where $\delta_{\pi p}$ is
negative, producing a sharp cutoff in the UHECR proton spectrum.

The empirical  constraint given by equation  (\ref{vc}) is independent
of  any constraint  on  $\delta_{\pi  p}$. However,  we  note that  if
$\delta_{\pi}  \simeq  \delta_p$, no  observable  modification of  the
UHECR spectrum occurs.  Therefore, we will assume that $\delta_{\pi} >
\delta_p$ at or near threshold  as a requirement for clearly observing
a potential LIV signal in the UHECR spectrum.  This assumption is also
made in Ref.  ~\cite{al03}.  We  will thus take $\delta_{\pi p} \equiv
\delta_{\pi}$ in  the case where  $\delta_p$ is small and  positive as
required by eq. (\ref{vc}).  Indeed, it can be shown in this case that
the dependence  of the  UHECR spectral shape  on the  $\delta_{\pi p}$
parameter dominates over that on the $\delta_p$ parameter \cite{ss08}.

\begin{figure}
\centerline{\psfig{figure=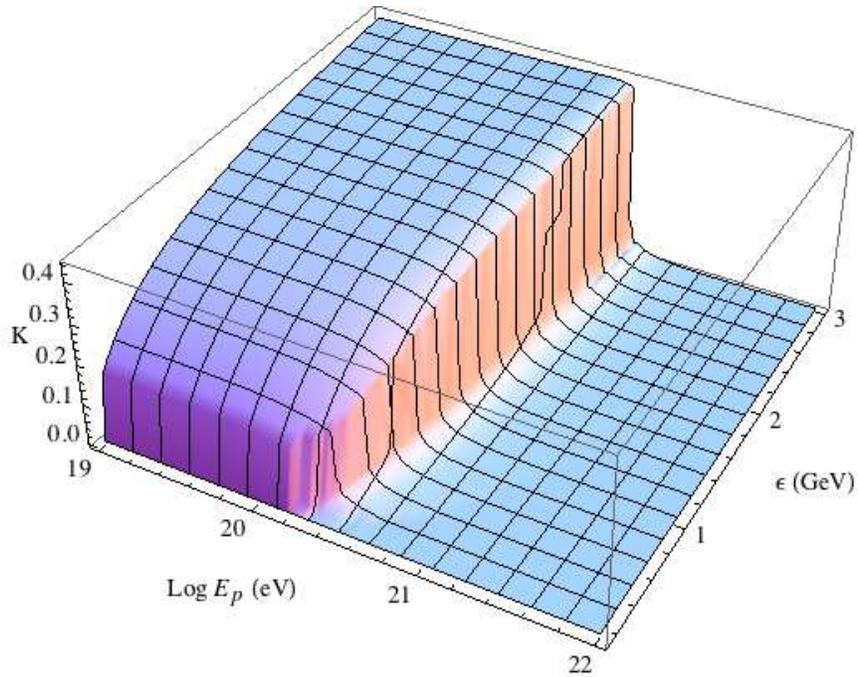,height=3.6in}}
\caption{The  calculated  proton  inelasticity  modified  by  LIV  for
$\delta_{\pi  p} =  3 \times  10^{-23}$ as  a function  of  CBR photon
energy and proton energy \protect \cite{ss08}.}
\label{inelasticity}
\end{figure}

Figure  \ref{inelasticity} shows  the  calculated proton  inelasticity
modified by LIV for a value of $\delta_{\pi p} = 3 \times 10^{-23}$ as
a function  of both CBR  photon energy and proton  energy \cite{ss08}.
Other choices for $\delta_{\pi p}$ yield similar plots.  The principal
result  of changing the  value of  $\delta_{\pi p}$  is to  change the
energy  at which  LIV effects  become  significant.  For  a choice  of
$\delta_{\pi p}  = 3 \times  10^{-23}$, there is no  observable effect
from LIV for $E_{p}$ less  than $\sim200$ EeV.  Above this energy, the
inelasticity  precipitously drops  as the  LIV term  in the  pion rest
energy approaches $m_{\pi}$.

With  this  modified inelasticity,  the  proton  energy  loss rate  by
photomeson production is given by

\begin{equation}
{{1}\over{E}}{{dE}\over{dt}}     =     -    {{\omega_{0}c}\over{2\pi^2
\gamma^2}\hbar^3c^3}  \int\limits_\eta^\infty d\epsilon  ~  \epsilon ~
\sigma(\epsilon)                                            K(\epsilon)
\ln[1-e^{-\epsilon/2\gamma\omega_{0}}]\end{equation}

\noindent where we  now use $\epsilon$ to designate  the energy of the
photon  in the  cms, $\eta$  is the  photon threshold  energy  for the
interaction in the cms, and $\sigma(\epsilon)$ is the total $\gamma$-p
cross  section   with  contributions  from   direct  pion  production,
multipion production, and the $\Delta$ resonance.

\begin{figure}
\centerline{\psfig{figure=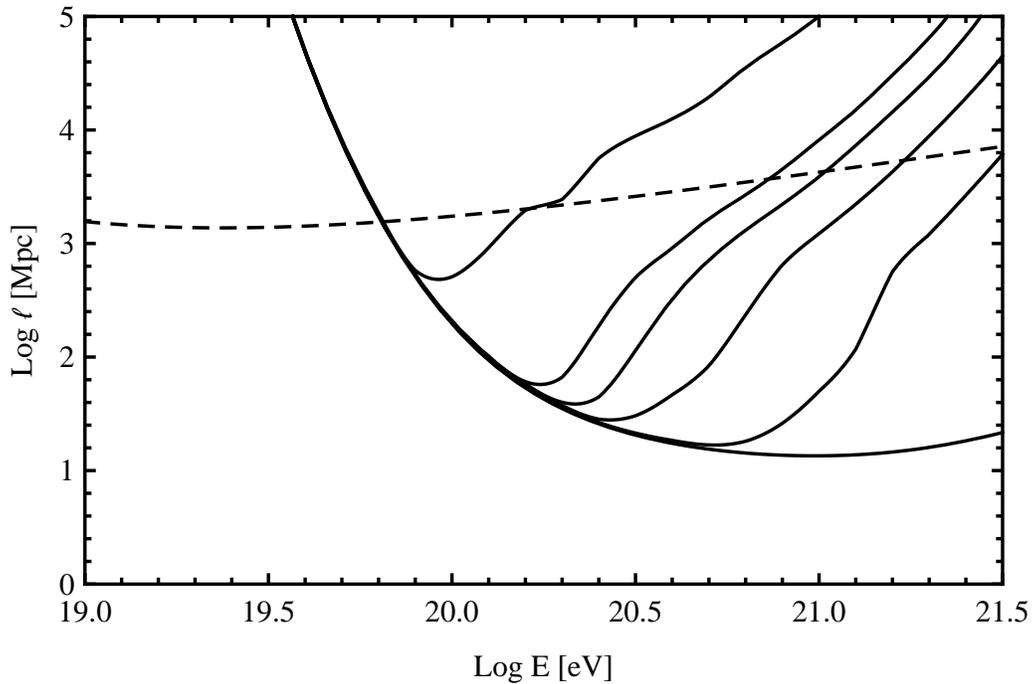,height=3.6in}}
\caption{The  calculated  proton  attenuation  lengths as  a  function
proton energy modified  by LIV for various values  of $\delta_{\pi p}$
(solid lines),  shown with the attenuation length  for pair production
unmodified by LIV  (dashed lines). From top to  bottom, the curves are
for $\delta_{\pi p} = 1 \times  10^{-22}, 3 \times 10^{-23} , 2 \times
10^{-23},  1  \times  10^{-23},  3  \times 10^{-24},  0$  (no  Lorentz
violation) \protect \cite{ss08}.}
\label{attn}
\end{figure}

The  corresponding  proton attenuation  length  is  given  by $\ell  =
cE/r(E)$,  where the  energy loss  rate $r(E)  \equiv  (dE/dt)$.  This
attenuation length is plotted  in Figure \ref{attn} for various values
of  $\delta_{\pi  p}$  along   with  the  unmodified  pair  production
attenuation   length   from  pair   production   interactions,  $p   +
\gamma_{CBR} \rightarrow e^+ + e^-$.

\section{UHECR Spectra with LIV and Comparison with Present Observations}

Let us now  calculate the modification of the  UHECR spectrum produced
by a very  small amount of LIV. We perform  an analytic calculation in
order  to determine the  shape of  the modified  spectrum.  It  can be
demonstrated that  there is little  difference between the  results of
using  an analytic  calculation {\it  vs.} a  Monte  Carlo calculation
({\it e.g.},  see Ref. \cite{ta09}). In  order to take  account of the
probable  redshift  evolution  of  UHECR  production  in  astronomical
sources, we take account of the following considerations: \\

\noindent  ({\it  i})  The  CBR  photon number  density  increases  as
$(1+z)^3$ and the CBR  photon energies increase linearly with $(1+z)$.
The corresponding energy loss for  protons at any redshift $z$ is thus
given by

\begin{eqnarray}
r_{\gamma p}(E,z) = (1+z)^3 r[(1+z)E].
\label{eq5}
\end{eqnarray}

\noindent  ({\it  ii})  We   assume  that  the  average  UHECR  volume
emissivity  is of  the energy  and  redshift dependent  form given  by
$q(E_i,z) =  K(z)E_i^{-\Gamma}$ where $E_i$  is the initial  energy of
the  proton  at  the source  and  $\Gamma  =  2.55$.  For  the  source
evolution, we assume $K(z) \propto (1  + z)^{3.6}$ with $z \le 2.5$ so
that  $K(z)$ is  roughly  proportional to  the empirically  determined
$z$-dependence of  the star formation rate. $K(z=0)$  and $\Gamma$ are
normalized fit the data below the GZK energy.\\

\begin{figure}
\centerline{\psfig{figure=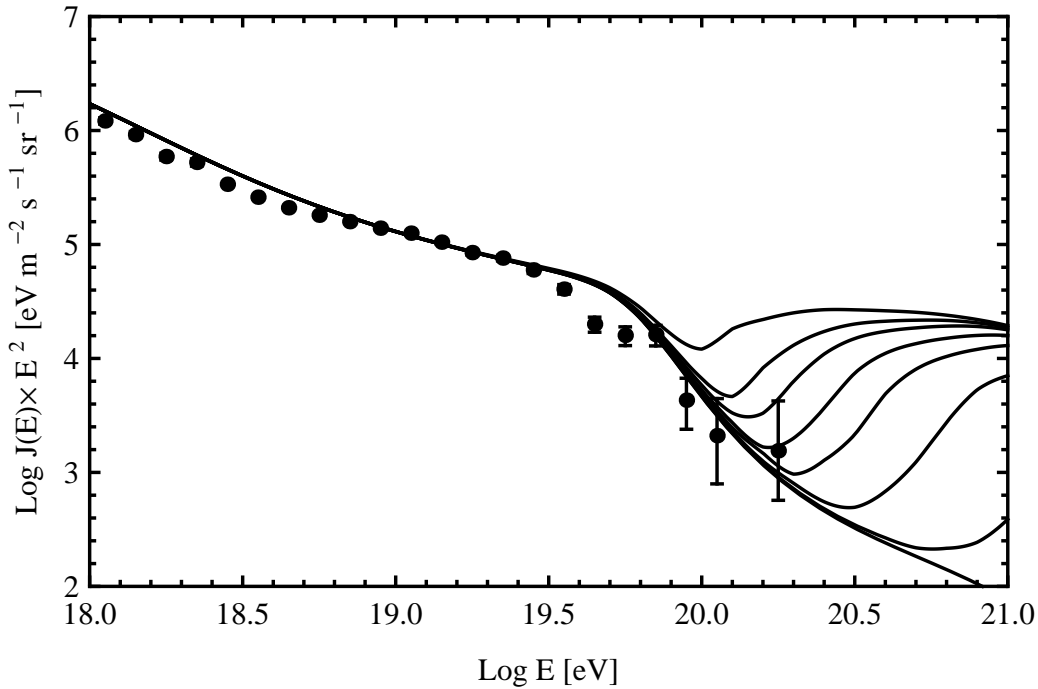,height=3.6in}}
\caption{Comparison of  the latest Auger data  with calculated spectra
for various  values of  $\delta_{\pi p}$, taking  $\delta_p =  0$ (see
text).  From top to bottom,  the curves give the predicted spectra for
$\delta_{\pi p}  = 1  \times 10^{-22}, 6  \times 10^{-23},  4.5 \times
10^{-23}, 3 \times 10^{-23} ,  2 \times 10^{-23}, 1 \times 10^{-23}, 3
\times 10^{-24}, 0$ (no Lorentz violation) \protect \cite{ss08}.}
\label{Auger}
\end{figure}

Using   these  assumptions,  Scully   and  Stecker   \cite{ss08}  have
calculated the effect  of LIV on the UHECR  spectrum.  The results are
actually  insensitive  to  the  assumed  redshift  dependence  because
evolution does not affect the shape of the UHECR spectrum near the GZK
cutoff  energy  ~\cite{be88,st05}.    At  higher  energies  where  the
attenuation length may again become  large owing to an LIV effect, the
effect  of evolution  turns  out to  be  less than  10\%.  The  curves
calculated in Ref. \cite{ss08} assuming various values of $\delta_{\pi
p}$, are shown  in Figure \ref{Auger} along with  the {\it Auger} data
from Ref. \cite{sch09}.  They show  that {\it even a very small amount
of LIV that is consistent with  both a GZK effect and with the present
UHECR data can lead to a  ``recovery'' of the UHECR spectrum at higher
energies.}

\begin{figure}
\centerline{\psfig{figure=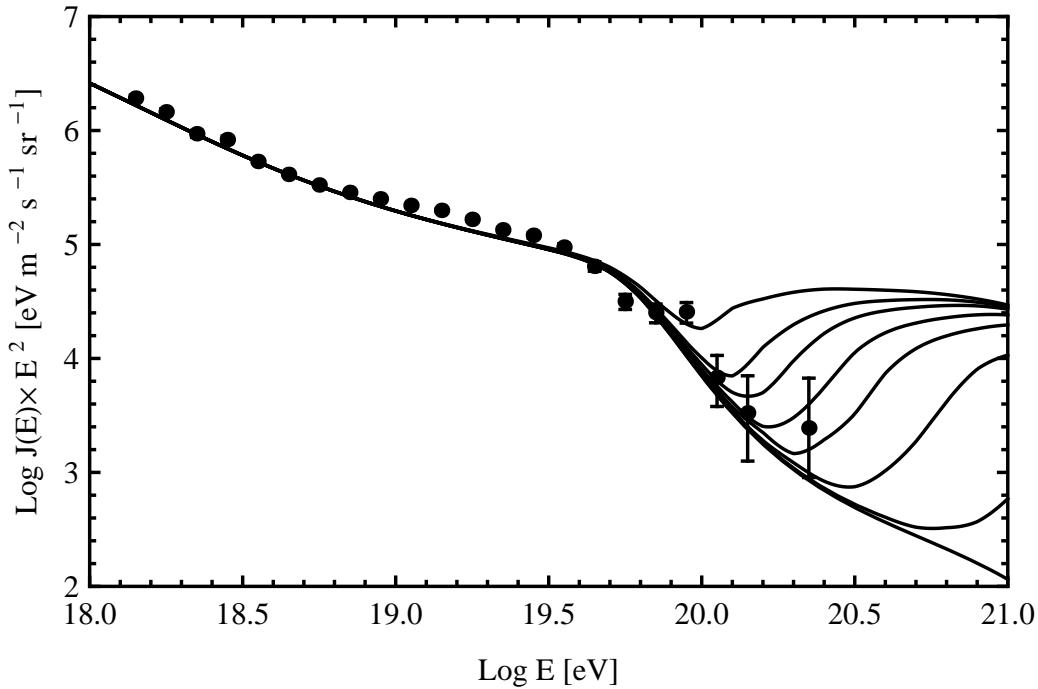,height=3.6in}}
\caption{Comparison  of the latest  Auger data  with the  stated UHECR
energies  increased  by 25\%  (see  text)  shown  with the  calculated
spectra for various values of $\delta_{\pi p}$, taking $\delta_p = 0$.
From  top  to  bottom,  the  curves give  the  predicted  spectra  for
$\delta_{\pi p}  = 1  \times 10^{-22}, 6  \times 10^{-23},  4.5 \times
10^{-23}, 3 \times 10^{-23} ,  2 \times 10^{-23}, 1 \times 10^{-23}, 3
\times 10^{-24}, 0$ (no Lorentz violation) \protect \cite{ss08}.}
\label{shift}
\end{figure}

\begin{figure}
\centerline{\psfig{figure=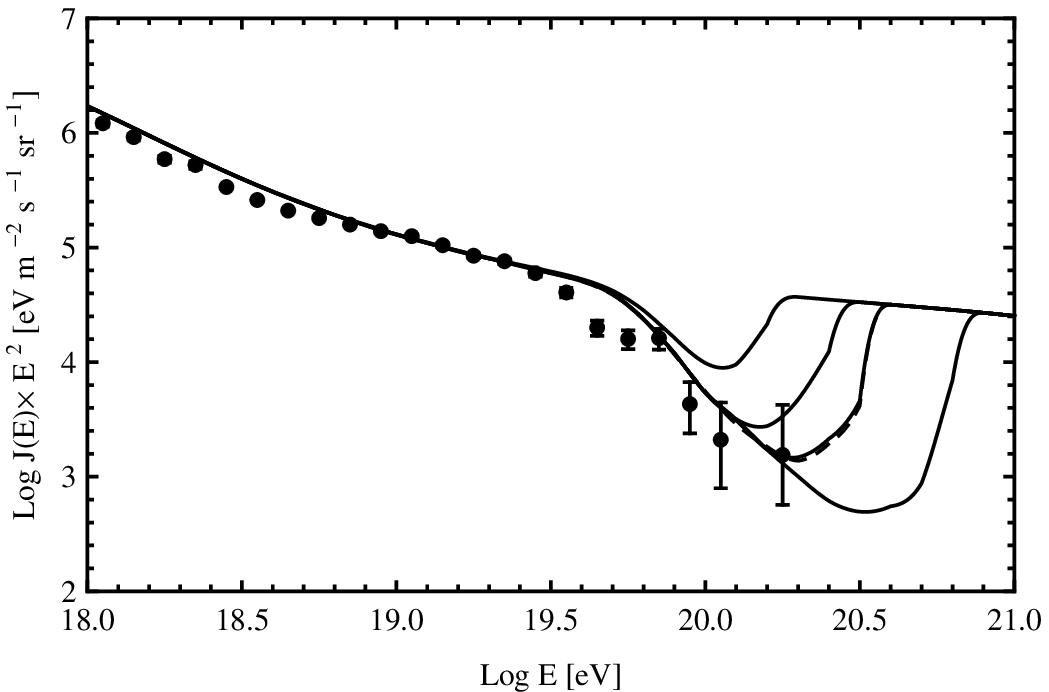,height=3.6in}}
\caption{Comparison of  the latest Auger data  with calculated spectra
for various  negative values  of $\delta_{p}$, taking  $\delta_{\pi} =
0$.  From  top to  bottom, the curves  give the predicted  spectra for
$\delta_{p}  = -  1 \times  10^{-23},  -5 \times  10^{-24}, -3  \times
10^{-24}, -1 \times 10^{-24}$. We also show by a dashed curve the case
where $\delta_p  = -3 \times  10^{-24}$ and $\delta_{\pi} =  -1 \times
10^{-24}$.}
\label{neg}
\end{figure}

\subsection{Non-Protonic UHECR}

Throughout this  paper, we have  made the assumption that  the highest
energy cosmic rays, {\it i.e.}, those above 100 EeV, are protons.  The
composition  of these  primary  particles is  presently unknown.   The
highest  energy events  for which  composition measurements  have been
attempted are in the range between  40 and 50 EeV, and the composition
of these events is uncertain~\cite{un09}-\cite{ul09}.

We note  that in  the case  where the UHECRs  with total  energy above
$\sim$100 EeV are  not protons, both the photomeson  threshold and the
LIV effects are moved to  higher energies because (i) the threshold is
dependent on $\gamma  \propto E/A$, where $A$ is  the atomic weight of
the   UHECR   \cite{st68},  and   (ii)   it   follows  from   equation
(\ref{dispersion})  that  the LIV  effect  depends  on the  individual
nucleon momentum

\begin{equation}
p_{N} \rightarrow E/A.
\end{equation}

In the case  of photodisintegration, LIV effects can  play a role.  We
note that  for single nucleon photodisintegration of  iron nuclei, the
threshold   is  at   a  higher   energy  than   for  the   GZK  effect
\cite{st69,st03a}.  For  He nuclei, on  the other hand,  the threshold
energy is lower than the GZK energy ~\cite{st69}.

\section{Constraints on LIV}

\subsection{Allowed Range for the LIV Parameter $\delta_{\pi p}$} 

It  has been  suggested  that  a small  amount  of Lorentz  invariance
violation (LIV)  could turn  off photomeson interactions  of ultrahigh
energy  cosmic rays  (UHECRs) with  photons of  the  cosmic background
radiation and thereby eliminate  the resulting sharp steepening in the
spectrum of the highest energy  CRs predicted by Greisen, Zatsepin and
Kuzmin (GZK).   Recent measurements of the UHECR  spectrum reported by
the   {\it   HiRes}   \cite{ab08}   and  {\it   Auger}   ~\cite{sch09}
collaborations,  however, appear  to indicate  the presence  of  a GZK
effect.

A true determination of  the implications of these recent measurements
for the search for  Lorentz invariance violation at ultrahigh energies
requires  a detailed  analysis of  the spectral  features  produced by
modifications  of  the  kinematical  relationships caused  by  LIV  at
ultrahigh energies. Scully and Stecker \cite{ss08} calculated modified
UHECR  spectra for  various values  of the  Coleman-Glashow parameter,
$\delta_{\pi  p}$,  defined  as  the difference  between  the  maximum
attainable velocities of the pion and the proton produced by LIV. They
then compared the results with the experimental UHECR data.

We have updated  these results using the very  latest {\it Auger} data
from the  procedings of the  2009 International Cosmic  Ray Conference
~\cite{sch09},\cite{data}. This update is shown in Figure \ref{Auger}.
The amount of presently observed  GZK suppression in the UHECR data is
consistent with the  possible existence of a small  amount of LIV.  In
order to  quantify this,  we determine the  value of  $\delta_{\pi p}$
that results in  the smallest $\chi^2$ for the  modeled UHECR spectral
fit using  the observational data from {\it  Auger} \cite{sch09} above
the GZK  energy.  The best  fit LIV parameter  found was in  the range
given  by  $\delta_{\pi  p}$  = $3.0^{+1.5}_{-3.0}  \times  10^{-23}$,
corresponding to  an upper  limit on $\delta_{\pi  p}$ of  $4.5 \times
10^{-23}$.
\footnote{The {\it HiRes} data \cite{ab08}  do not reach a high enough
energy  to further  restrict  LIV.}   This result,  as  it stands,  is
slightly     more     constraining      than     that     given     in
Ref.~\cite{ss08}. However, we note that the overall fit of the data to
the theoretically expected spectrum  is somewhat imperfect, even below
the GZK energy  and even for the  case of no LIV. It  appears that the
{\it Auger} spectrum seems to steepen even  below the GZK energy. 
As a conjecture,
we have taken  the liberty of assuming that the  derived energy may be
too low by about 25\%,  within the uncertainty of both systematic-plus
statistical error  given for  the energy determination.  By increasing
the derived  UHECR energies by  25\%, we arrive  at the plot  shown in
Figure \ref{shift}, again shown with the theoretical curves. In Figure
\ref{shift} one  sees better agreement between  the theoretical curves
and the  shifted data.  The constraint on  LIV would be  only slightly
reduced if this shift is assumed.

The results for  LIV modified spectra given in  Ref.  \cite{ss08} were
calculated under the assumption that $\delta_p \equiv \delta_{p\gamma}
= 0$.  It follows from equation (\ref{restmass}) that if $\delta_p$ is
slightly negative  then the spectra are  additionally modified because
of the  reduced cms  energy of  the proton for  a given  lab momentum.
This  affects  the photomeson  interaction  rate  in  a different  and
stronger way  than for the $\delta_{p\gamma}  = 0$ case  shown in Fig.
\ref{Auger}.   Here,  the  reduced  cms  proton energy  results  in  a
reduction  of  the  phase  space  allowed for  the  interactions  when
$\sqrt{s_p}$ given by equation ({\ref{restmass}) is near zero.

The   results   for   negative   $\delta_p$  are   shown   in   Figure
\ref{neg}. They  lie within the  range of constraints  on $\delta_{\pi
p}$ given  above.  However, it is  clear that even  a relatively small
negative value for  $\delta_p$ has a stronger LIV  effect on the UHECR
spectrum than a positive value  of $\delta_{\pi}$. We also show a case
where $\delta_{p}$  and $\delta_{\pi}$ are both  negative (dashed line
in the  figure).  The  dashed curve shows  that the  same $\delta_{p}$
produces an  almost identical  effect on the  spectrum in  both cases,
again demonstrating that the negative $\delta_{p}$ parameter gives the
dominant LIV effect.

We  also present  here, for  comparison, the  spectrum for  a slightly
positive  $\delta_{p\gamma}$. Figure  \ref{vcr-spec} shows  two curves
for $\delta_{\pi p} = 5  \times 10^{-23}$.  The spectrum with a vacuum
\v{C}erenkov radiation  cutoff at 300  EeV is for  $\delta_{p\gamma} =
0.5  \times  10^{-23}$ (see  equation  (\ref{vc})).   The other  curve
assumes $\delta_{p\gamma} = 0$ as in Figure \ref{Auger}.

\begin{figure}
  \centerline{\psfig{figure=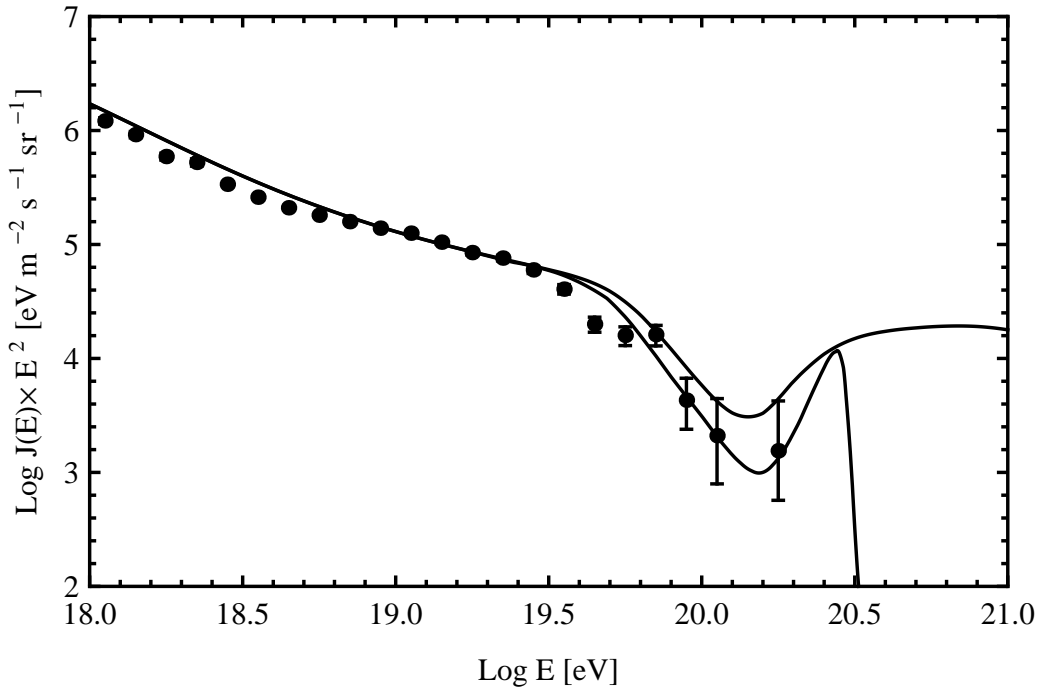,height=3.6in}}
  \caption{Comparison of the latest Auger data with calculated spectra
for $\delta_{\pi p} = 4.5 \times 10^{-23}$ and for $\delta_p = 0 $ and
$0.5 \times 10^{-23}$ as discussed in the text.  In the later case the
cutoff from the vacuum \v{C}erenkov effect is apparent.}
\label{vcr-spec}
\end{figure}

\subsection{Implications for Quantum Gravity Models}

An  effective  field theory  approximation  for  possible LIV  effects
induced by Planck-scale suppressed  quantum gravity for $E \ll M_{Pl}$
was considered  in Ref. \cite{ma09}.  These authors  explored the case
where a  perturbation to  the energy-momentum dispersion  relation for
free particles would be produced  by a CPT-even dimension six operator
suppressed  by a  term  proportional to  $M_{Pl}^{-2}$. The  resulting
dispersion relation for a particle of type $a$ is

\begin{equation}
E_a^2 = p_a^2 + m_a^2 + \eta_a \left( {{p^4}\over{M_{Pl}^2}} \right)
\label{QG}
\end{equation} 

In order  to explore the  implications of our constraints  for quantum
gravity,  we will  equate  the perturbative  terms  in the  dispersion
relation given  by our equation (\ref{dispersiona}),  for both protons
and  pions, with  the  equivalent dimension  six dispersion  relations
given by  equation (\ref{QG}). We  note that the perturbative  term in
equation (\ref{QG})  has an  energy dependence, whereas  our dimension
four case does  not.  However, since we are  only comparing with UHECR
data over  a very limited energy  range around a  fiducial energy $E_f
\sim$ 100 EeV, we will make the identification at that energy.

Using  this  identification,  we  find  that  in  most  cases  an  LIV
constraint  of $\delta_{\pi  p} <  4.5  \times 10^{-23}$  at a  proton
fiducial energy  of $E_f \sim  100$ EeV indirectly implies  a powerful
limit on the representation of quantum gravity effects in an effective
field   theory  formalism   with  Planck   suppressed   dimension  six
operators. Equating the perturbative terms in both the proton and pion
dispersion relations

\begin{equation}
2\delta_{\pi     p}    \simeq     (\eta_{\pi}    -     25    \eta_{p})
\left({{0.2E_f}\over{M_{Pl}}}\right)^2 ,
\label{dim6}
\end{equation}

\noindent where  we have adopted  the terminology of  Ref. \cite{ma09}
and we have taken the pion fiducial energy to be $\sim 0.2 E_f$, as at
the $\Delta$ resonance \cite{st68}.   Since we require $\delta_{\pi p}
> 0$  for GZK suppression,  and $\delta_p  > 0$  in order  to suppress
proton  vacuum  \v{C}erenkov  radiation,  equation  (\ref{dim6})  then
implies that  $\delta_{\pi} > \delta_p$, which is  the assumption made
in  Refs.  \cite{al03}  and \cite{ss08}.   Equation  (\ref{dim6}) also
indicates  that LIV  by dimension  six  operators is  suppressed by  a
factor  of at  least ${\cal{O}}(10^{-6}  M_{Pl}^{-2})$, except  in the
unlikely  case   that  $\eta_{\pi}-  25  \eta_{p}   \simeq  0$.   This
suppression is over and above that  of any dimension four terms in the
dispersion relation as we have  considered here.  These results are in
agreement with  the conclusions of  Ref.  \cite{ma09} who also  find a
suppression  of  ${\cal{O}}(10^{-6}  M_{Pl}^{-2})$,  except  with  the
equivalent loophole.   We note that  in Ref.  \cite{ma09} a  series of
Monte Carlo  runs are used in  order to obtain their  results.  It can
thus  be concluded that  an effective  field theory  representation of
quantum gravity  with dimension six  operators that suppresses  LIV by
only  a factor of  $M_{Pl}^2$ is  effectively ruled  out by  the UHECR
observations, as concluded in Ref. \cite{ma09}.

\section{Beyond Constraints: Seeking LIV}

As we have seen (see Figure  \ref{Auger}), even a very small amount of
LIV that  is consistent with  both a GZK  effect and with  the present
UHECR data can lead to a ``recovery'' of the primary UHECR spectrum at
higher energies. This is the  clearest and the most sensitive evidence
of an LIV signature. The  ``recovery'' effect has also been deduced in
Refs.~\cite{ma09} and ~\cite{bi09}
\footnote{In Ref.~\cite{bi09},  a recovery effect is  also claimed for
high proton energies in the  case when $\delta_{\pi p} < 0$.  However,
we have noted that  the `quasi-vacuum \v{C}erenkov radiation' of pions
by  protons in  this case  will  cut off  the proton  spectrum and  no
``recovery'' effect will  occur.}. In order to find  it (if it exists)
three conditions must exist: ({\it i}) sensitive enough detectors need
to be built, ({\it ii}) a  primary UHECR spectrum that extends to high
enough energies ($\sim$ 1000 EeV) must exist, and ({\it iii}) one much
be able to distinguish the LIV signature from other possible effects.

\subsection{Cosmic Zevatrons}

In  order to meet  our second condition, we  require the  existence of
powerful cosmic ray accelerators,  the so-called zevatrons (1000 EeV =
1  ZeV)  .  In  this  ``bottom  up''  scenario for  UHECR  production,
acceleration in  extragalactic sources must account  for the existence
of  observed UHECRs  with  energies reaching  at  least $\sim$300  EeV
~\cite{bi94}.   The most widely  considered acceleration  mechanism is
shock  acceleration,  particularly  in  the lobes  of  powerful  radio
galaxies ({\it  e.g.}, Refs.  ~\cite{ca78}  -- ~\cite{de09}.) Blanford
~\cite{bl99}   discusses    the   problems   associated    with   this
``conventional''  mechanism of accelerating  particles to  the highest
observed energies.  Other  acceleration mechanisms have been proposed.
In particular, it  has recently been argued that  the plasma wakefield
acceleration  mechanism, operating  within relativistic  AGN  jets, is
capable   of  accelerating   particles  to   energies  $\sim$   1  ZeV
~\cite{ch09}.  If  such zevatrons exist  and produce particles  of ZeV
energies, given enough statistics from future detector studies, an LIV
signal can be searched for.

\subsection{Distinguishing an LIV Signal}

\subsubsection{LIV {\it vs} the  Top-Down Scenario:}

Our signature  signal of  LIV is a  ``recovery'' of the  primary UHECR
spectrum  at higher energies  (see Figure  \ref{Auger}).  Such  an LIV
signal  must be  distinguished from  the presence  of a  higher energy
component in the UHECR spectrum  predicted to be produced by so-called
``top-down''  models.   The top-down  scenarios  invoke  the decay  or
annihilation  of  supermassive  particles  or topological  or  quantum
remnants of the very early universe usually associated with some grand
unification energy scale.  Such  processes result in the production of
a high ratio of pions to nucleons from the resulting QCD fragmentation
process (see , {\it e.g.},  ~\cite{st03a} for a review).  Owing to QCD
fragmentation, top-down  scenarios predict relatively  large fluxes of
UHE  photons and  neutrinos  as compared  to  nucleons, as  well as  a
significant diffuse GeV background flux  that could be searched for by
the {\it Fermi} $\gamma$-ray space telescope.

A  higher energy  UHECR  component arising  from  top-down models  can
indeed  be  distinguished  from  the  LIV  effect.   Contrary  to  the
predictions  of relatively  copious  pion production  in the  top-down
scenario, the  LIV effect cuts off  UHE pion production  at the higher
energies and  consequent UHE neutrino  and photon production  from UHE
pion decay.  We  also note that LIV would therefore  not produce a GeV
photon flux.

In   this  regard,  we   note  that   the  Pierre   Auger  Observatory
collaboration  has  provided observational  upper  limits  on the  UHE
photon  flux \cite{au08,au09}  that have  already  disfavored top-down
models.  The upper limits from the {\it Auger} array indicate that UHE
photons  make up  at best  only a  small percentage  of the  total UHE
flux. This contradicts predictions of top-down models that the flux of
UHE photons should be larger than that of UHE protons. Upper limits on
the UHE neutrino flux from {\it ANITA} also strongly disfavor top-down
models \cite{anita06}, \cite{go08}.

\subsubsection{LIV {\it vs.} Local Source Overdensity:}

It is possible that the  apparent modified GZK suppression in the data
may be related to an overdensity  of nearby sources related to a local
supergalactic enhancement~\cite{st68}. However, at this point in time,
no  clear correlation  of UHECR  directions with  nearby extragalactic
sources exists.\footnote{A  correlation with nearby AGN  was hinted at
in  earlier Auger  data~\cite{2ab08}.   However, the  HiRes group  has
found no  significant correlation~\cite{abb08} and  no correlation has
now been found in the more  recent Auger data with an increased number
of events (Westerhoff, private  communication).}  More and better data
will be required in order to resolve this question.  An LIV effect can
be distinguished from a  posssible local source enhancement by looking
for UHECRs at energies above $\sim$200 EeV, as can be seen from Figure
\ref{Auger}.  This  is because the small  amount of LIV  that fits the
observational UHECR spectra can lead  to the signature recovery of the
cosmic ray  flux at higher  energies than presently observed.   Such a
recovery is not expected in the case of a local overdensity. Searching
for such  a recovery effect will  require obtaining a  future data set
containing a much higher number of UHECR air shower events.

\section{Obtaining UHECR Data at Higher Energies}

We now  turn to examining the  various techniques that can  be used in
the  future  in  order  to  look  for a  signal  of  LIV  using  UHECR
observations.   As   can  be  seen  from   the  preceding  discussion,
observations of higher energy  UHECRs with much better statistics than
presently obtained  are needed in order  to search for  the effects of
miniscule Lorentz invariance violation on the UHECR spectrum.

\subsection{Auger North}

In the  future, such  an increased number  of events may  be obtained.
The {\it  Auger} collaboration has  proposed to build an  ``{\it Auger
North''}  array that  would be  seven  times larger  than the  present
southern hemisphere Auger array ({\tt http://www.augernorth.org}).

\subsection{Future Space Based Detectors}

Further  into  the future,  space-based  telescopes  designed to  look
downward  at large  areas of  the  Earth's atmosphere  as a  sensitive
detector  system  for giant  air-showers  caused  by trans-GZK  cosmic
rays. We  look forward to  these developments that may  have important
implications for fundamental high energy physics.

Two  future  potential spaced-based  missions  have  been proposed  to
extend  our knowledge  of  UHECRs  to higher  energies.   One is  {\it
JEM-EUSO}  (the Extreme  Universe Space  Observatory)  ~\cite{EUSO}, a
one-satellite telescope mission proposed  to be placed on the Japanese
Experiment Module (JEM) on the International Space Station.  The other
is {\it OWL} (Orbiting Wide-angle Light Collectors) ~\cite{OWL}, a two
satellite mission for stereo viewing, proposed for a future free-flyer
mission.   Such  orbiting  space-based  telescopes with  UV  sensitive
cameras will have  wide fields-of-view (FOVs) in order  to observe and
use large  volumes of  the Earth's atmosphere  as a  detecting medium.
They will thus trace the atmospheric fluorescence trails of numbers of
giant  air  showers  produced  by  ultrahigh energy  cosmic  rays  and
neutrinos.   Their large  FOVs will  allow the  detection of  the rare
giant air  showers with energies higher than  those presently observed
by  ground-based detectors such  as {\it  Auger}.  Such  missions will
thus  potentially open  up  a new  window  on physics  at the  highest
possible observed energies.

\subsubsection{The Extreme Universe Space Observatory {\it JEM-EUSO}:}

{\it JEM-EUSO} has been selected  as one of two candidate missions for
the  second  utilization of  the  Japanese  Experiment  Module on  the
International Space Station. It may  be launched in 2013 by a Japanese
heavy lift rocket. It will employ a double plastic Fresnel lens system
telescope  with  a  30$^\circ$  FOV  and  will  help  to  advance  the
technology of such missions.  Further into the future, a proposed {\it
``Super EUSO''} mission is in the preliminary planning stage.

\subsubsection{The {\it OWL} Mission Concept:}

The {\it OWL} (Orbiting  Wide-field Light-collectors), a proposed dual
satellite mission to have a larger total aperture than {\it JEM-EUSO},
has  been designed to  be sensitive  enough to  obtain data  on higher
energy  UHECRs  and on  ultrahigh  energy  neutrinos.   Its
detecting  area and  FOV will  be large  enough to  provide  the event
statistics and  extended energy range  that are crucial  to addressing
these issues.  To accomplish this, {\it OWL} will also make use of the
Earth's  atmosphere  as a  huge  ``calorimeter"  to make  stereoscopic
measurements of the atmospheric UV fluorescence produced by air shower
particles.   {\it  OWL} is  thus  proposed to  consist  of  a pair  of
satellites  placed in  tandem in  a low  inclination,  medium altitude
orbit. The {\it OWL} telescopes will  point down at the Earth and will
together point at a section of  atmosphere about the size of the state
of Texas ($\sim 6 \times 10^5$ km$^2$).

The baseline {\it OWL}  instrument, shown in Figure \ref{OWL1} (left),
is a  large f/1 Schmidt  camera with a  $45^{\circ}$ full FOV and  a 3
meter entrance aperture.  The entrance aperture will contain a Schmidt
corrector. The deployable primary mirror has a 7 meter diameter.  {\it
OWL} would be normally operated in stereo mode and the two ``OWL eye''
instruments will view a common volume of atmosphere.

\begin{figure}[h]
\begin{center}
\mbox{\psfig{figure=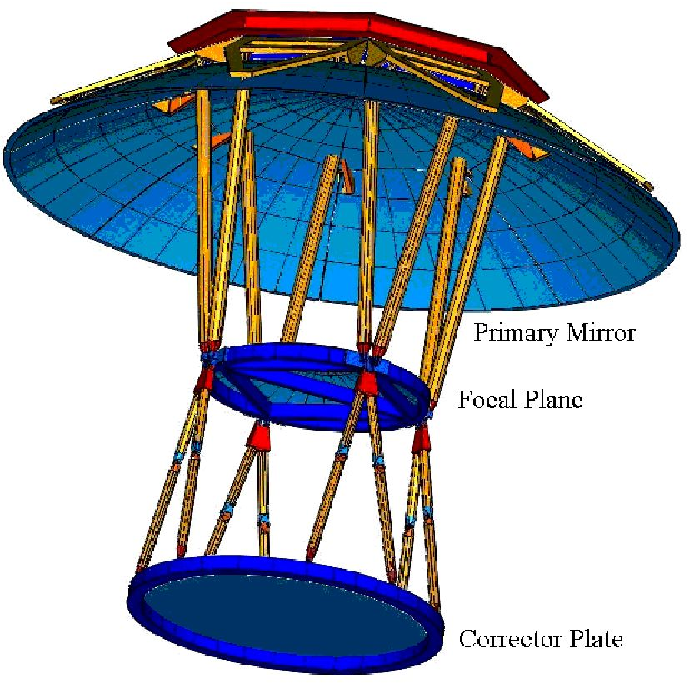,height=9cm}}
\hspace{1.cm}
\mbox{\psfig{figure=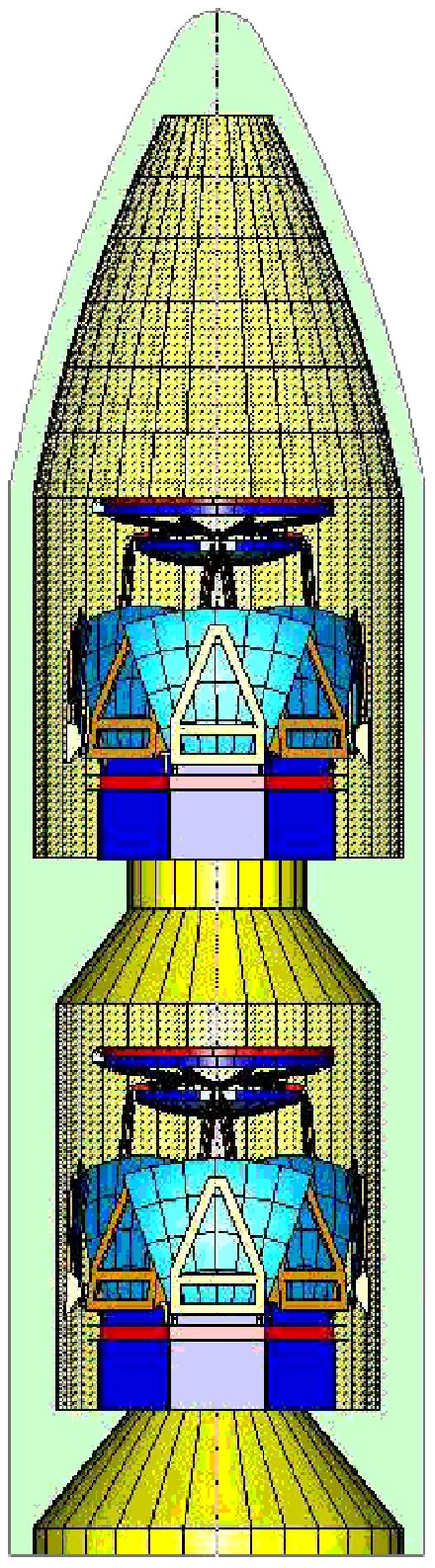,height=10cm}}
\end{center}
\caption{Left: Schematic of the Schmidt  optics that form an {\it OWL}
``eye''  in the  deployed  configuration.  The  spacecraft bus,  light
shield, and  shutter are  not shown.  Right:  Schematic of  the stowed
{\it OWL} satellites in the launch vehicle.}
\label{OWL1}
\end{figure}

The  satellites can  be launched  together on  a Delta  rocket  into a
proposed 1000  km circular orbit with an  inclination of $10^{\circ}$.
Figure \ref{OWL1}  (right) shows both satellites stowed  for launch as
well as  a depiction of  one of the Schmidt  telescopes.  Stereoscopic
observation resolves  spatial ambiguities and  allows determination of
corrections  for  the  effects  of  clouds.  In  stereo,  fast  timing
provides supplementary  information to reduce  systematics and improve
the resolution of the arrival direction of the UHECR. By using stereo,
differences in  atmospheric absorption or  scattering of the  UV light
can be  determined. Detector missions  such as the proposed  {\it OWL}
mission  can provide  the statistics  of  UHECR events  that would  be
needed in the  100 to 1000 EeV energy range to  search for the effects
of a very small amount  of Lorentz invariance violation at the highest
energies.

\begin{figure}
\centerline{\psfig{figure=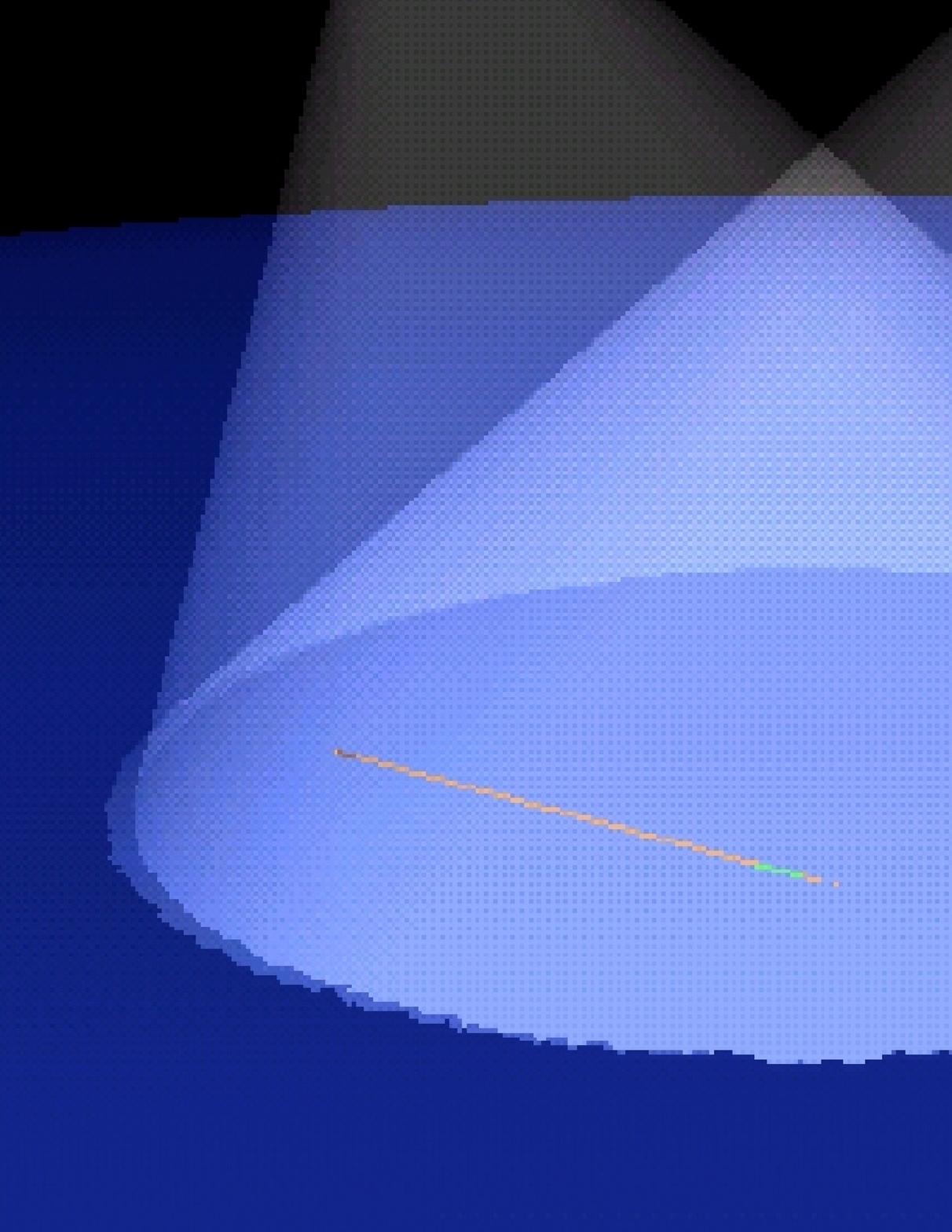,height=8cm}}

\caption{Two   OWL  satellites  in   low-Earth  orbit   observing  the
fluorescent track of  a giant air shower. The  shaded cones illustrate
the field-of-view for each satellite.}
\label{OWL2}
\end{figure}

We look forward to such future detector
developments. As we have seen, they may have important implications 
for fundamental high energy physics as well as the astrophysics
of powerful extragalactic ``zevatrons''.

\section*{Acknowledgments}

We would like to thank John F. Krizmanic for helpful discussions.
We would also like to thank Alan Watson for bringing the new {\it Auger}
data to our attention.

\end{document}